\documentclass[11pt]{article}
\usepackage{moriond,epsfig}

\def\be{\begin{equation}}
\def\ee{\end{equation}}
\def\bea{\begin{eqnarray}}
\def\eea{\end{eqnarray}}

\def\dsup{D_{\rm sup}}
\def\dfav{D_{\rm fav}}
\def\dpisig{6.4\sigma}
\def\dksig{2.3\sigma}
\def\rb{r_B}
\def\rblimit{0.27}
\def\rd{r_D}
\def\deltab{\delta_B}
\def\deltad{\delta_D}
\def\stat{{\rm stat}}
\def\syst{{\rm syst}}

\newcommand{\bdk}{$B^{\pm}\to DK^{\pm}$}
\newcommand{\bdtk}{$B^{\pm}\to \tilde{D}K^{\pm}$}

\newcommand{\bdsk}{$B^{\pm}\to D^{*}K^{\pm}$}
\newcommand{\bdstk}{$B^{\pm}\to \tilde{D}^{*}K^{\pm}$}

\newcommand{\bddsk}{$B^{\pm}\to D^{(*)}K^{\pm}$}

\begin{document}

\newcommand{\de}{\Delta E}
\newcommand{\mbc}{M_{\rm bc}}
\newcommand{\bb}{B{\bar B}}
\newcommand{\qq}{q{\bar q}}

\vspace*{4cm}
\title{\Large Measurement of $\phi_3$}

\author{ P. Krokovny}

\address{High Energy Accelerator Research
  Organization (KEK), Tsukuba, Japan}

\maketitle\abstracts{
We present results related to determination 
of the Unitarity Triangle angle $\phi_3$.
}
\section{Introduction}
\begin{figure}
  \begin{center}
  \epsfig{figure=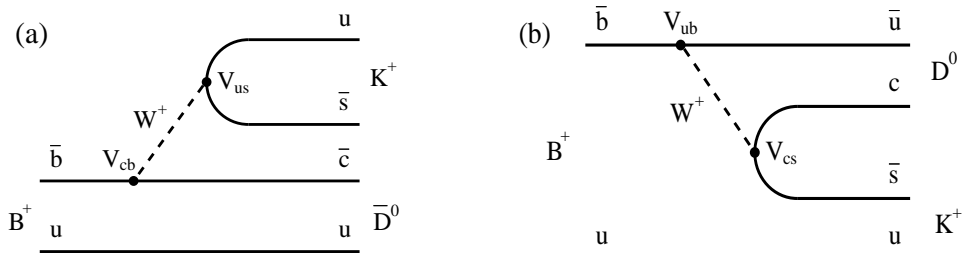,width=0.8\textwidth}
  \caption{Feynman diagrams of (a) dominant $B^+\to \bar{D^0}K^+$ and 
           (b) suppressed $B^+\to D^0K^+$ decays}
  \label{diags}
  \end{center}
\end{figure}

Determinations of the Cabibbo-Kobayashi-Maskawa
(CKM) \cite{ckm} matrix elements provide important checks on
the consistency of the Standard Model and ways to search
for new physics.  Various methods using $CP$ violation in $B\to D K$ 
decays have been proposed~\cite{glw,dunietz,eilam,ads,giri} to 
measure the Unitarity Triangle
angle $\phi_3$. These methods are based on two key observations:
neutral $D^{0}$ and $\bar{D^0}$
mesons can decay to a common final state, and the decay
$B^+\to D^{(*)} K^+$ can produce neutral $D$ mesons of both flavors
via $\bar{b}\to \bar{c}u\bar{s}$ (Fig.~\ref{diags}a)
and $\bar{b}\to \bar{u}c\bar{s}$ (Fig.~\ref{diags}b) transitions,
with a relative phase $\theta_+$ between the two interfering
amplitudes that is the sum, $\delta + \phi_3$, of strong and weak 
interaction phases.  For the charge conjugate mode, the relative phase 
is $\theta_-=\delta-\phi_3$. 

The results are based on a data sample containing 275 million
$B\bar{B}$ pairs, 
collected with the Belle detector~\cite{NIM} at the KEKB asymmetric energy $e^+e^-$ 
collider operating at the $\Upsilon(4S)$ resonance.

\section{$B^\pm \rightarrow D_{CP}K^\pm$ and $D^{*}_{CP} K^\pm$}
Recent theoretical studies on $B$ meson dynamics have
demonstrated a method to access $\phi_3$ using the process $B^-\rightarrow
D^{(*)0}K^{(*)-}$~\cite{glw}. When a $D^0$ is reconstructed as a $CP$ eigenstate, the
$b\rightarrow c$ and $b\rightarrow u$ processes interfere. 
This interference leads to direct $CP$ violation. 
To extract
$\phi_3$ and assuming no $D^0-\bar{D}^0$ mixing, some necessary
observables sensitive to $CP$ violation are:

\begin{eqnarray*}
{\cal{A}}_{1,2} & \equiv & \frac{{\cal B}(B^- \rightarrow D_{1,2}K^-) -
{\cal B}(B^+ \rightarrow D_{1,2}K^+) }{{\cal B}(B^- \rightarrow
D_{1,2}K^-) + {\cal B}(B^+ \rightarrow D_{1,2}K^+) }
 = \frac{2 r \sin \delta ' \sin \phi_3}{1 + r^2 + 2 r \cos \delta '
\cos \phi_3}\\
{\cal{R}}_{1,2} & \equiv & \frac{R^{D_{1,2}}}{R^{D^{0}}}  = 1 + r^2 + 2 r
\cos \delta ' \cos \phi_3,\,\,\,\,\,\,\,\,\,\,\,\,\,\,
\delta '  =  \left\{
             \begin{array}{ll}
              \delta & \mbox {{\rm  for }$D_1$}\\
              \delta + \pi&  \mbox{{\rm for }$D_2$}\\
             \end{array}
             \right.,
\end{eqnarray*}

where the ratios $R^{D_{1,2}}$ and $R^{D^{0}}$ are defined as
\begin{eqnarray*}
R^{D_{1,2}} & = & \frac{{\cal B}(B^- \rightarrow D_{1,2}K^-)+{\cal B}(B^+
\rightarrow D_{1,2}K^+)}{{\cal B}(B^- \rightarrow D_{1,2}\pi^-) +
{\cal B}(B^+ \rightarrow D_{1,2}\pi^+)}\\
        R^{D^{0}} & = & \frac{{\cal
B}(B^- \rightarrow D^{0} K^-)+{\cal B}(B^+ \rightarrow
\bar{D}^{0} K^+)}{{\cal B}(B^- \rightarrow D^{0}\pi^-) + {\cal B}(B^+
\rightarrow \bar{D}^{0} \pi^+)}
\end{eqnarray*}

\noindent where $D_1$ and $D_2$ are $CP$-even and
$CP$-odd eigenstates respectively. The asymmetries ${\cal{A}}_{1}$ and
${\cal{A}}_{2}$ have opposite signs. The ratio $r$ is defined as 
$r = |A(B^-\to\bar{D}^0 K^-)/A(B^- \to D^0 K^-)|$ and is the ratio of 
the two tree diagrams shown in Fig.~\ref{diags} where 
$\delta$ is their strong-phase difference. 
The size of the ratio $r$ governs the magnitude of the maximum possible
CP asymmetry; this ratio is suppressed by both CKM $(\sim0.45)$ and color
$(\sim 0.40)$ factors.
The asymmetries and double ratios can be
calculated for $D^*$ in a similar manner.
The analysis is described in detail elsewhere~\cite{belle_dcp}.

\begin{figure}
\begin{center}
\begin{tabular}{cc}
\includegraphics[width=0.3\textwidth]{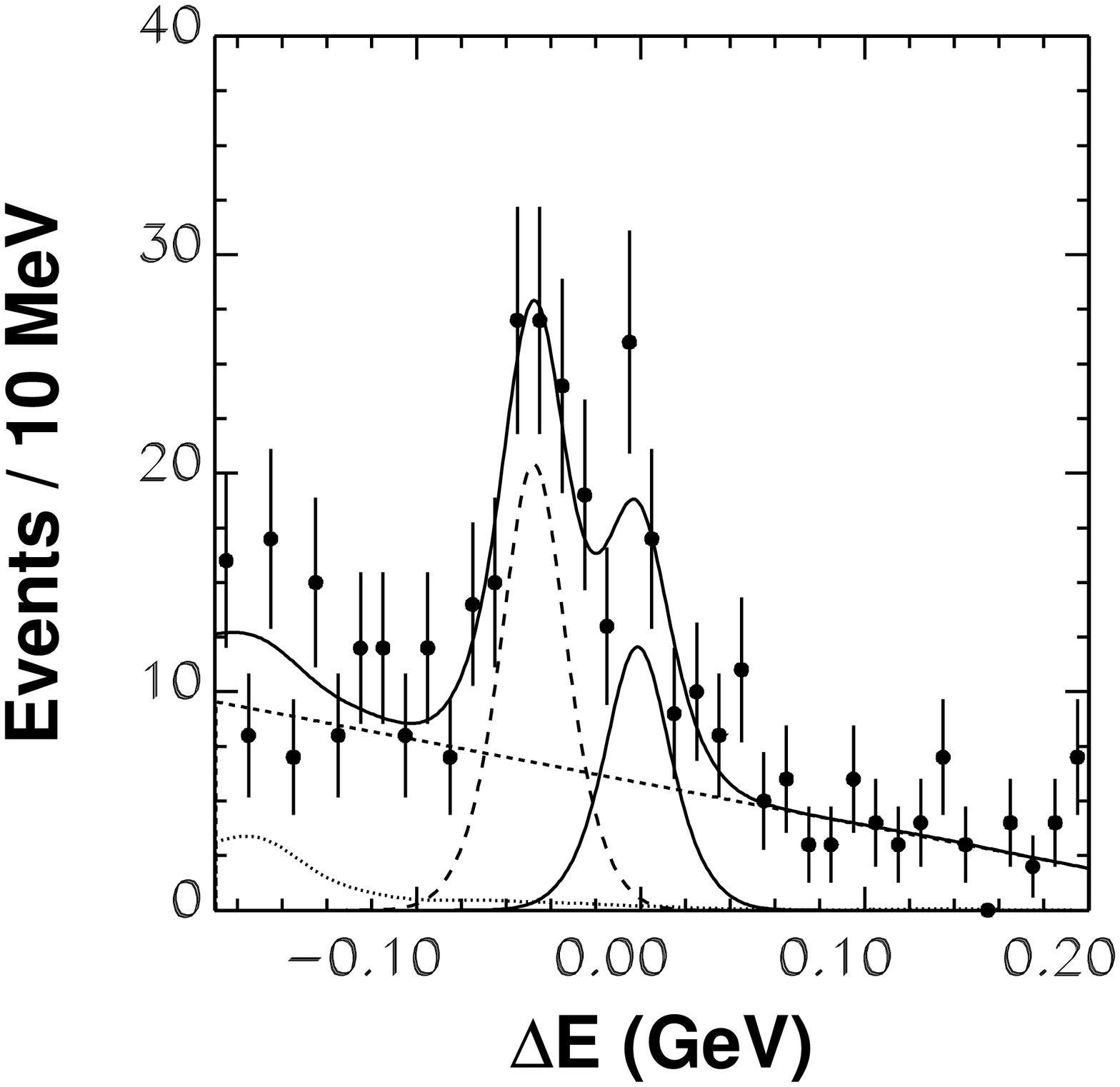} &
\includegraphics[width=0.3\textwidth]{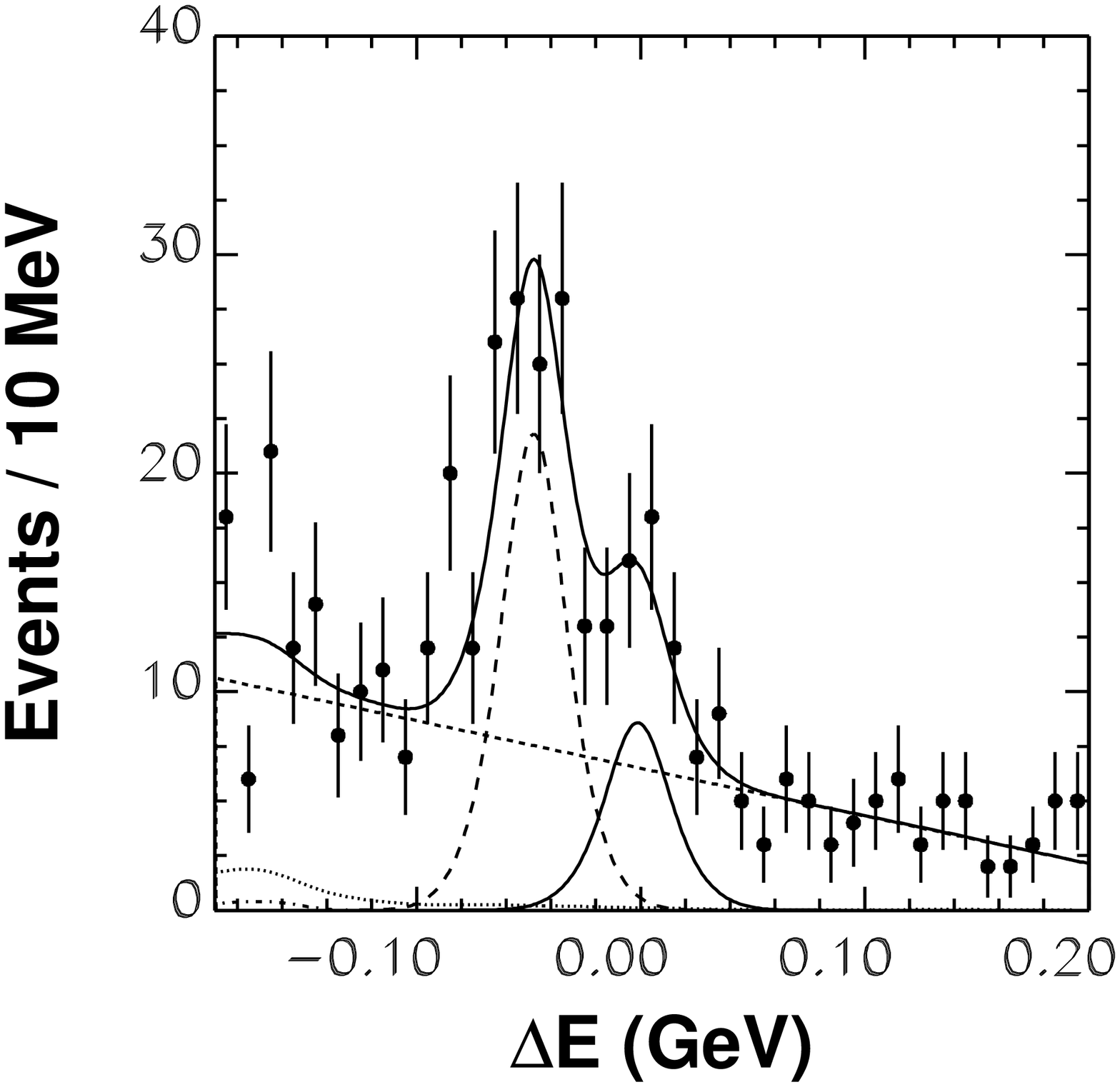} \\
\includegraphics[width=0.3\textwidth]{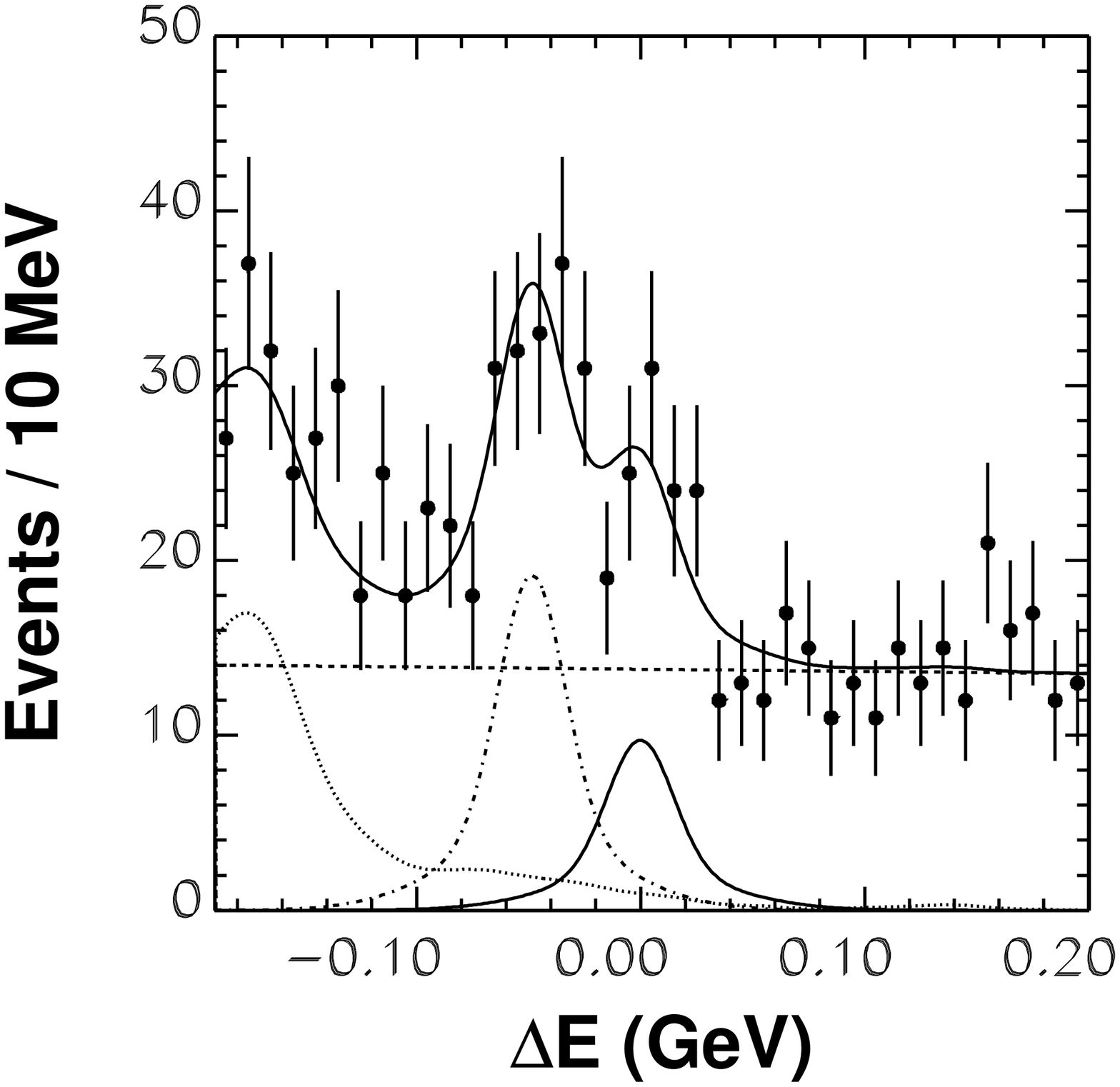} &
\includegraphics[width=0.3\textwidth]{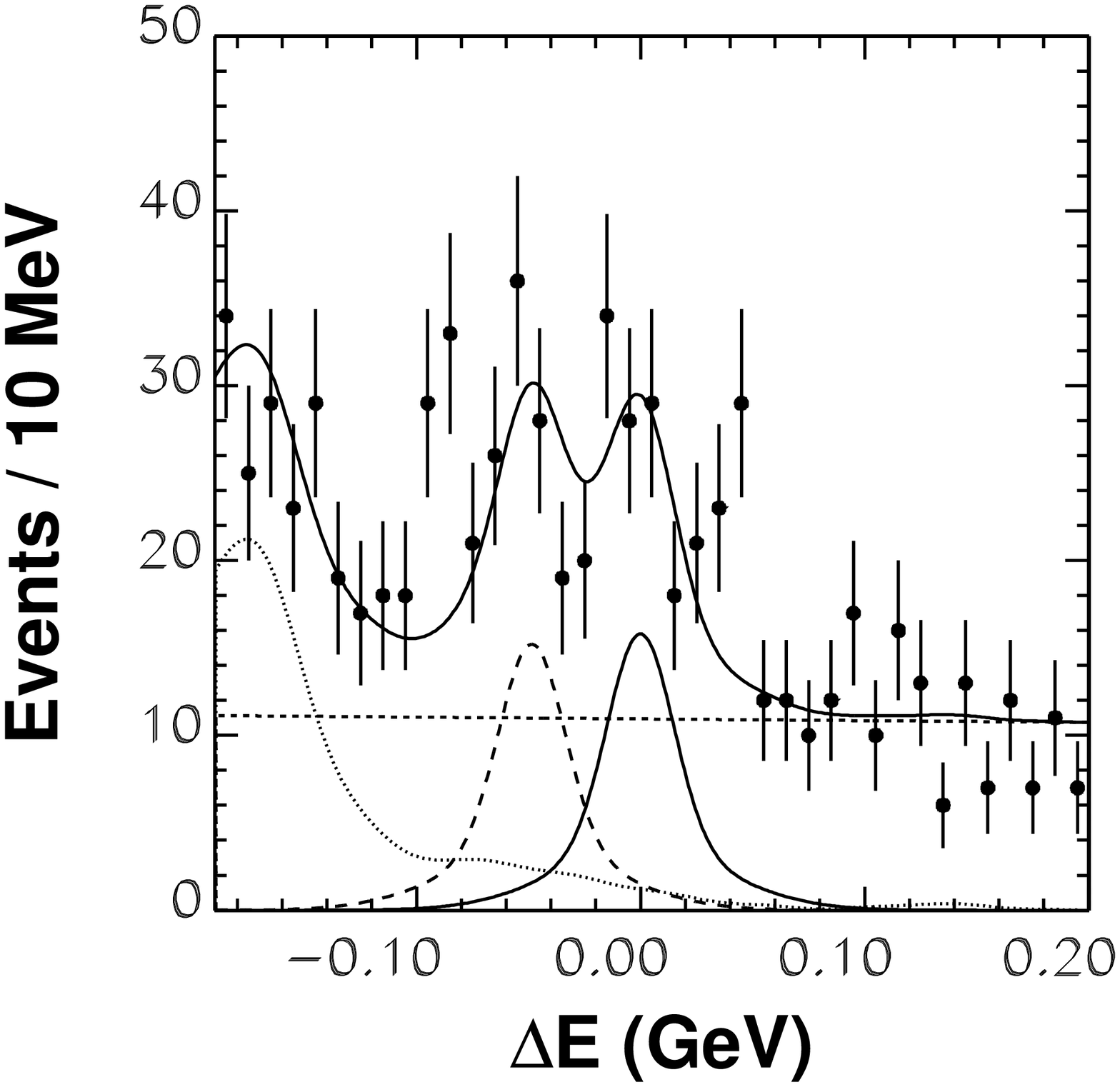} \\
\end{tabular}
\end{center}
\caption{$\Delta E$ distributions for (top left) $B^+ \rightarrow D_1 K^+$, (top right) $B^-\rightarrow
D_1 K^-$, (bottom left) $B^+ \rightarrow
D_2 K^+$, (bottom right) $B^- \rightarrow
D_2 K^-$.}
\label{dk2}
\end{figure}

Fig.~\ref{dk2} shows the $\Delta E$ distributions for $B^\pm\to D
K^\pm$ events. 
Table~\ref{table:asym_dh} summarizes the yields from $\Delta E$ fit 
and the corresponding asymmetries with statistical errors. 
In the control samples, no large deviation from 0 is seen. The modes of 
interest are $D_1K$ and $D_2K$ where the $B^+$ and $B^-$ events are 
used to calculate asymmetries and double ratios. 

\begin{table*}
\caption{Yields and asymmetries obtained for $Dh$ modes. }
\begin{tabular*}{\textwidth}{l@{\extracolsep{\fill}}cccc}
\hline\hline
 &  $\sum B$  & $B^{+}$  & $B^{-}$ & $\cal A$\\
\hline $D_f\pi$ & 19283 $\pm$ 150 & 9690 $\pm$ 104 &
9521 $\pm$ 103 & -0.01 $\pm$ 0.01\\
$D_1\pi$ & 2183 $\pm$ 55 & 1051 $\pm$ 36 &
1132 $\pm$ 37 & 0.04  $\pm$ 0.02\\
$D_2\pi$ & 2413 $\pm$ 93 & 1178 $\pm$ 43 &
1226 $\pm$ 43 & 0.02 $\pm$ 0.03 \\
$D_fK$ & 1031 $\pm$ 39 & 484 $\pm$ 27 &
549 $\pm$ 28 & 0.06 $\pm$ 0.04\\
$D_1K$ & 114 $\pm$ 21 & 49 $\pm$ 15 &
63 $\pm$ 14 & 0.07 $\pm$ 0.14 \\
$D_2K$ & 167 $\pm$ 21 & 94 $\pm$ 17 &
75 $\pm$ 16 & -0.11 $\pm$ 0.14 \\
\hline\hline
\end{tabular*}
\label{table:asym_dh}
\end{table*}



\par The final asymmetries for ${\cal A}_{1}$ and ${\cal A}_{2}$ are 
found do be

\begin{center}
\begin{tabular}{lcc}
${\cal A}_1$ & = & 0.07 $\pm$ 0.14 (stat) $\pm$ 0.06 (sys) \\
${\cal A}_2$ & = & -0.11 $\pm$ 0.14 (stat) $\pm$ 0.05 (sys) \\
\end{tabular}
\end{center}
agreeing with theoretical expectations where they should have
opposite signs. The double ratios:
\begin{center}
\begin{tabular}{lcc}
$R_1$ & = & 0.98 $\pm$ 0.18 (stat) $\pm$ 0.10 (sys) \\
$R_2$ & = & 1.29 $\pm$ 0.16 (stat) $\pm$ 0.08 (sys) \\
\end{tabular}
\end{center}

\begin{figure}
\begin{center}
\begin{tabular}{cc}
\includegraphics[width=0.3\textwidth]{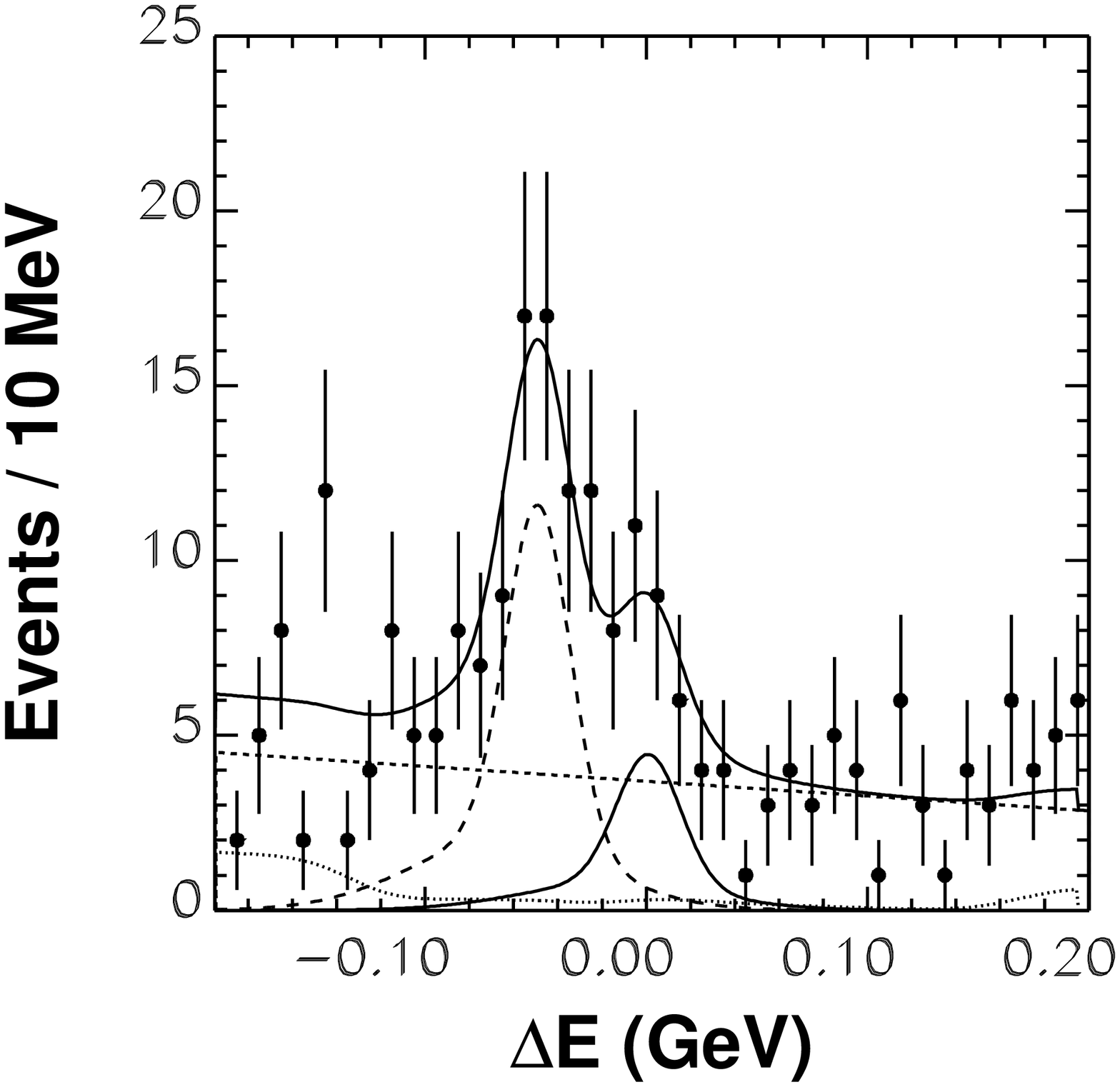}
\includegraphics[width=0.3\textwidth]{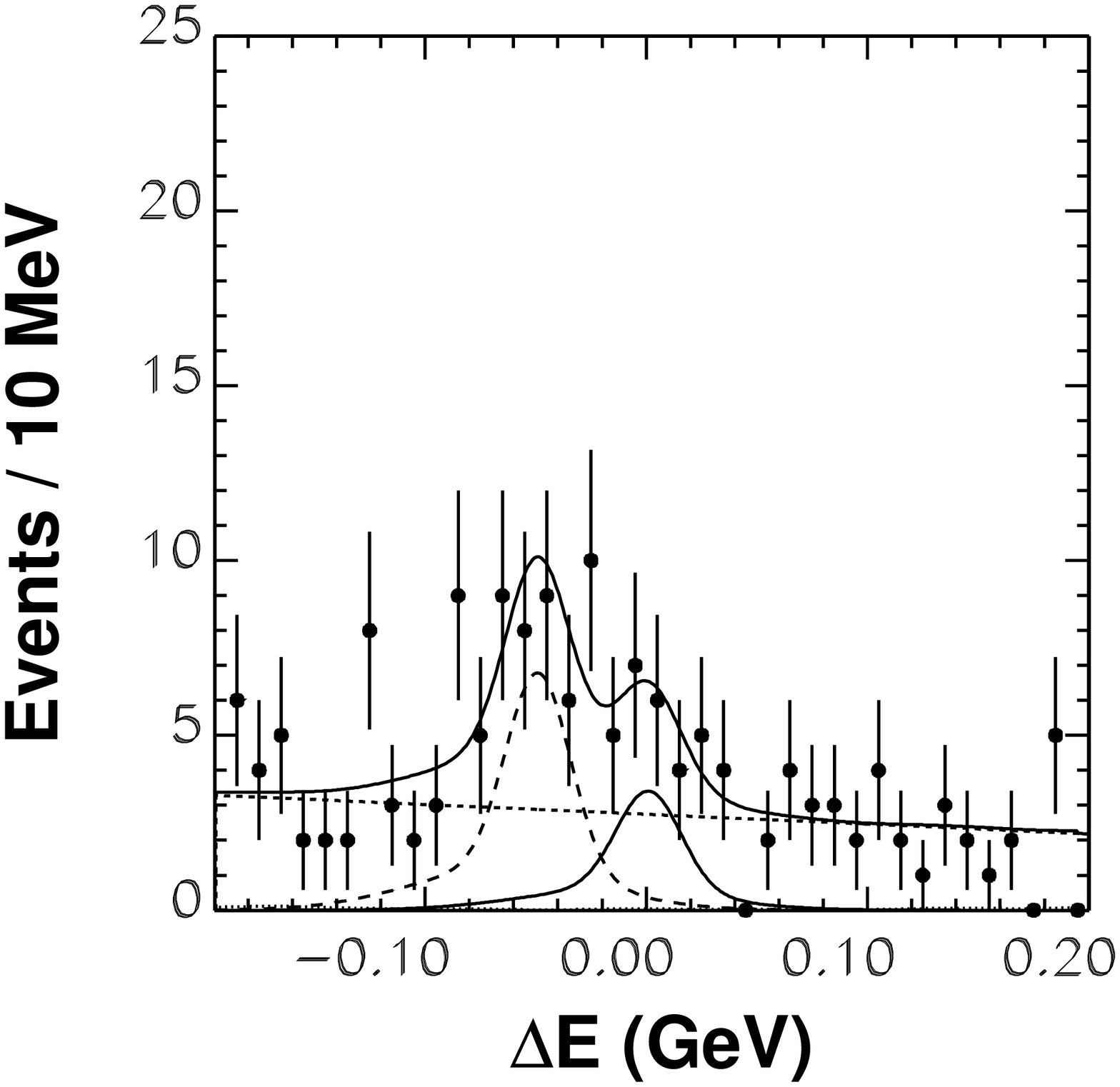}
\end{tabular}
\end{center}
\caption{$\Delta E$ distributions for (left) $B^\pm \rightarrow
D^{*0}_1K^\pm$, (right) $B^\pm \rightarrow
D^{*0}_2K^\pm$.}
\label{dstk}
\end{figure}

Fig.~\ref{dstk} shows the $\Delta E$ distributions for $B^\pm\to D^*
K^\pm$ events. 
Table~\ref{table:dst_asym} contains the yields of the
distributions with statistical errors and asymmetries. 
The statistical significance of $D^*_{1} K$ and $D^*_{2} K$ signals
are 5.6 and 4.5 respectively.

\begin{table*}
\caption{Yields and asymmetries obtained for $D^*h$ modes.}
\begin{tabular*}{\textwidth}{l@{\extracolsep{\fill}}cccc}
\hline\hline
 &  $\sum B$  & $B^{+}$  & $B^{-}$ & $\cal A$\\
\hline
$D^*\pi$ & 5762 $\pm$ 101 & 2681 $\pm$ 74 & 2594 $\pm$ 74 & -0.02 $\pm$ 0.02\\
$D^*_{1}\pi$ &  795 $\pm$ 41 & 399 $\pm$ 24 & 397 $\pm$ 23 & 0.00 $\pm$ 0.04 \\
$D^*_{2}\pi$ &  715 $\pm$ 37 & 415 $\pm$ 35 & 301 $\pm$ 33 & -0.16 $\pm$ 0.07\\
$D^* K$ &  284 $\pm$ 23 & 158 $\pm$ 16 & 127 $\pm$ 16 & -0.11 $\pm$ 0.08\\
$D^*_{1} K$ & 56 $\pm$ 11 & 33 $\pm$ 8 & 19 $\pm$ 8 & -0.27 $\pm$ 0.25 \\
$D^*_{2} K$ & 33 $\pm$ 10 & 13 $\pm$ 6 & 22 $\pm$ 7 & 0.26 $\pm$ 0.26 \\
\hline\hline
\end{tabular*}
\label{table:dst_asym}
\end{table*}

Asymmetries were found to be:
\begin{center}
\begin{tabular}{lcc}
${\cal A}_1$ & = & -0.27 $\pm$ 0.25 (stat) $\pm$ 0.04 (sys) \\
${\cal A}_2$ & = & 0.26 $\pm$ 0.26 (stat) $\pm$ 0.03 (sys) \\
\end{tabular}
\end{center}
where the systematic errors were calculated in a similar way to the
$Dh$ case. Double ratios found are:
\begin{center}
\begin{tabular}{lcc}
$R_1$ & = & 1.43 $\pm$ 0.28 (stat) $\pm$ 0.06 (sys) \\
$R_2$ & = & 0.94 $\pm$ 0.28 (stat) $\pm$ 0.06 (sys)
\end{tabular}
\end{center}

In summary, the partial rate asymmetries ${\cal A}_{1,2}$ are
measured for the decays $B^\pm \rightarrow D^{(*)}_{CP}K^\pm$
and are consistent with zero.
A first observation is seen for $D^*_1 K$ and $D^*_2 K$.

\section{Measurement of $\phi_3$ with Dalitz Plot
Analysis of \boldmath{\bddsk} Decay}
Recently, three body final states common to $D^0$ and
$\bar{D^0}$, such as $K_S\pi^+\pi^-$ \cite{giri}, were suggested as 
promising modes for the extraction of $\phi_3$. 
This method is based on two key observations:  neutral $D^{0}$ and $\bar{D}^{0}$
mesons can decay to a common final state such as $K_s \pi^+ \pi^-$, 
and the decay
$B^+\to D^{(*)} K^+$ can produce neutral $D$ mesons of both flavors 
via $\bar{b}\to \bar{c}u\bar{s}$ and $\bar{b}\to \bar{u}c\bar{s}$ 
transitions, where the relative phase $\theta_+$ between the two interfering
amplitudes is the sum, $\delta + \phi_3$, of strong and weak interaction
phases.  In the charge conjugate mode, the relative phase
$\theta_-=\delta-\phi_3$, so both phases can be extracted 
from the measurements of such $B$ decays and their charge conjugate modes. 
The phase measurement is based
on the analysis of Dalitz distribution of the three body final state of the 
$D^{0}$ meson.
The analysis is described in detail 
elsewhere~\cite{belle_dalitz}.



The Dalitz plots of $D$ decaying to $K_s\pi^+\pi^-$, 
which contain information about CP violation in $B$
decays, are fitted for $B^-$ and $B^+$ data sets.
A combined unbinned maximum likelihood fit to the 
$B^+$ and $B^-$ samples with 
$r$, $\phi_3$ and $\delta$ as free parameters yields the following values: 
$r=0.25\pm 0.07$, $\phi_3=64^{\circ}\pm 15^{\circ}$, 
$\delta=157^{\circ}\pm 16^{\circ}$ for the \bdtk\ sample and 
$r=0.25\pm 0.12$, $\phi_3=75^{\circ}\pm 25^{\circ}$, 
$\delta=321^{\circ}\pm 25^{\circ}$ for the \bdstk\ sample.
The errors quoted here are obtained from the likelihood fit.
These errors are a good representation of the statistical uncertainties for
a Gaussian likelihood distribution, however in our case
the distributions are highly non-Gaussian. In addition, the errors
for the strong and weak phases depend on the values of the
amplitude ratio $r$ ({\it e.g.} for $r=0$ there is 
no sensitivity to the phases). A more reliable estimate of the
statistical uncertainties is obtained using a large number
of MC pseudo-experiments as discussed below.

We use a frequentist technique to evaluate the 
statistical significance of the measurements. 
To obtain the  probability density function (PDF) of the fitted 
parameters as a function of the true parameters, which is needed for this 
method, we employ a ``toy" MC technique that uses a
simplified MC simulation of the experiment which incorporates
the same efficiencies, resolution and backgrounds as
used in the data fit.  This MC is used
to generate several hundred experiments for a given set of
$r$, $\theta_+$ and $\theta_-$ values. For each simulated
experiment, Dalitz plot distributions are generated
with equal numbers of events as in the data, 137 and 139 events 
for $B^-$ and $B^+$ decays, correspondingly, for \bdtk\ mode and 
34 and 35 events for $B^-$ and $B^+$ for
\bdstk\ mode. The simulated Dalitz plot distributions
are subjected to the same fitting procedure that is applied
to the data. This is repeated for different values of $r$,
producing distributions of the fitted parameters that
are used to produce a functional form of the PDFs of the 
reconstructed values for any set of input parameters.

\begin{figure}
\hspace{3\baselineskip}
  \epsfig{figure=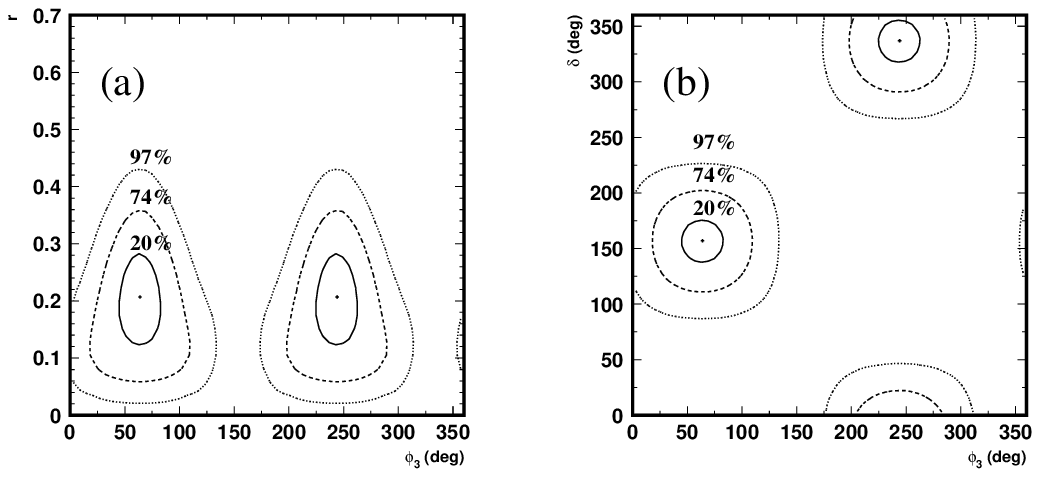,width=0.8\textwidth}
  \caption{Confidence regions for the pairs of parameters (a) ($r$, $\phi_3$) 
           and (b) ($\phi_3, \delta$) for the \bdtk\ sample.}
  \label{b2dk_neum}
\end{figure}

\begin{figure}
\hspace{3\baselineskip}
  \epsfig{figure=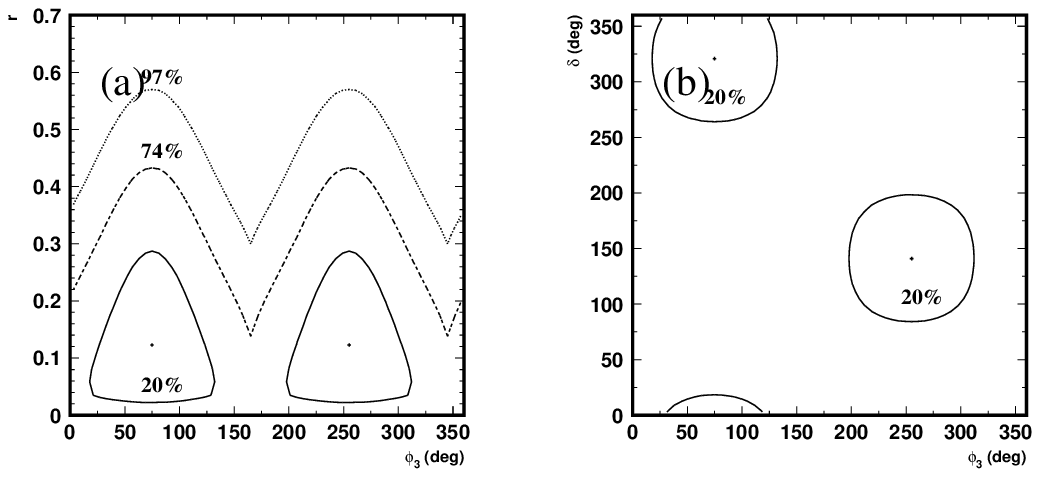,width=0.8\textwidth}
  \caption{Confidence regions for the pairs of parameters (a) ($r$, $\phi_3$)
           and (b) ($\phi_3, \delta$) for the \bdstk\ sample.}
  \label{b2dsk_neum}
\end{figure}

The confidence regions for the pairs of parameters 
$(\phi_3, \delta)$ and $(\phi_3, r)$ are shown in Fig.~\ref{b2dk_neum} 
(\bdtk\ mode) and Fig.~\ref{b2dsk_neum} (\bdstk\ mode).
They are the projections of the corresponding 
confidence regions in the three-dimensional parameter space. 
We show the 20\%, 74\% and 97\% confidence level regions, 
which correspond to 
one, two, and three standard deviations for a three-dimensional Gaussian
distribution.

For the final results, we use the central values that  are obtained by 
maximizing the PDF and the statistical errors corresponding to the 20\% 
confidence region (one standard deviation). Of the two possible 
solutions ($\phi_3$, $\delta$ and $\phi_3+180^{\circ}$, $\delta+180^{\circ}$) 
we choose the one with $0<\phi_3<180^{\circ}$. The final results are 
\begin{equation} 
r = 0.21 \pm 0.08 \pm 0.03 \pm 0.04,
~\phi_3=64^{\circ} \pm 19^{\circ} \pm 13^{\circ} \pm 11^{\circ},~ 
\delta=157^{\circ} \pm 19^{\circ} \pm 11^{\circ} \pm 21^{\circ}
\end{equation}
for the \bdtk\ mode and 
\begin{equation}
r = 0.12^{+0.16}_{-0.11} \pm 0.02 \pm 0.04,
~\phi_3=75^{\circ} \pm 57^{\circ} \pm 11^{\circ} \pm 11^{\circ},
~\delta=321^{\circ} \pm 57^{\circ} \pm 11^{\circ} \pm 21^{\circ}
\end{equation}
for the \bdstk\ mode.  The first, second, and third errors are
statistical, systematic, and model dependent errors.

The significance of $CP$ violation is 94\% for the \bdtk\ sample
and 38\% for \bdstk.

The two events samples, \bdk\ and \bdsk, are combined 
in order to obtain a more accurate measurement of $\phi_3$. 
The $\phi_3$ result from the combined analysis is 
\begin{equation}
\phi_3=68^{\circ}\;^{+14^{\circ}}_{-15^{\circ}}\pm 13^{\circ}\pm 11^{\circ},
\end{equation}
where the first error is statistical, the second is experimental systematics, and
the third is model uncertainty. 
The two standard deviation interval including the 
systematic and model uncertainties is $22^{\circ}<\phi_3<113^{\circ}$. 
The statistical significance of $CP$ violation for the combined measurement 
is 98\%. 

\section{Study of the Suppressed Decays 
  $B^- \to [K^+\pi^-]_D K^-$ and $B^- \to [K^+\pi^-]_D \pi^-$}

As noted by Atwood, Dunietz and Soni (ADS)~\cite{ads},
$CP$ violation effects are enhanced if the final state is chosen so
that the interfering amplitudes have comparable magnitudes;
the archetype uses $B^- \to [K^+\pi^-]_D K^-$,
where $[K^+\pi^-]_{D}$ indicates that the $K^+\pi^-$ pair 
originates from a neutral $D$ meson. 
The analysis is described in detail elsewhere~\cite{belle_ads}.

\begin{table*}
  \caption{
    Summary of the results. 
    For the $B^- \to \dsup K^-$ signal yield, 
    the peaking background contribution has been subtracted. 
    The first two errors on the measured production branching fractions
    are statistical and systematic, respectively,
    and the third is due to the uncertainty in the $B^- \to \dfav h^-$
    product branching fraction used for normalization.
  }
\begin{tabular*}{\textwidth}{l@{\extracolsep{\fill}}c c c c}
\hline\hline
      Mode & Signal Yield & Statistical  & Measured product & Upper limit \\
           & & significance  & branching fraction & ($90\%$C.L.)\\
      \hline
      $B^- \to \dsup K^-$ & 
      $   8.5\,^{ +6.0}_{ -5.3}$ & $\dksig$  & $(3.2\,^{+2.2}_{-2.0} \pm 0.2 \pm 0.5) \times 10^{-7}$ & $6.3 \times 10^{-7}$ \\
      $B^- \to \dsup \pi^-$ & 
      $  28.5\,^{ +8.1}_{ -7.4}$ & $\dpisig$ & $(6.6\,^{+1.9}_{-1.7} \pm 0.4 \pm 0.3) \times 10^{-7}$ & $-$ \\
      $B^- \to \dfav K^-$ & 
      $ 376.0\,^{+21.8}_{-21.1}$ & $-$       & $-$ & $-$ \\
      $B^- \to \dfav \pi^-$ & 
      $8181.9\,^{+94.0}_{-93.3}$ & $-$       & $-$ & $-$ \\
\hline\hline
    \end{tabular*}
  \label{tab:yield}
\end{table*}

\begin{figure}
  \begin{center}
    \includegraphics[width=0.5\textwidth]{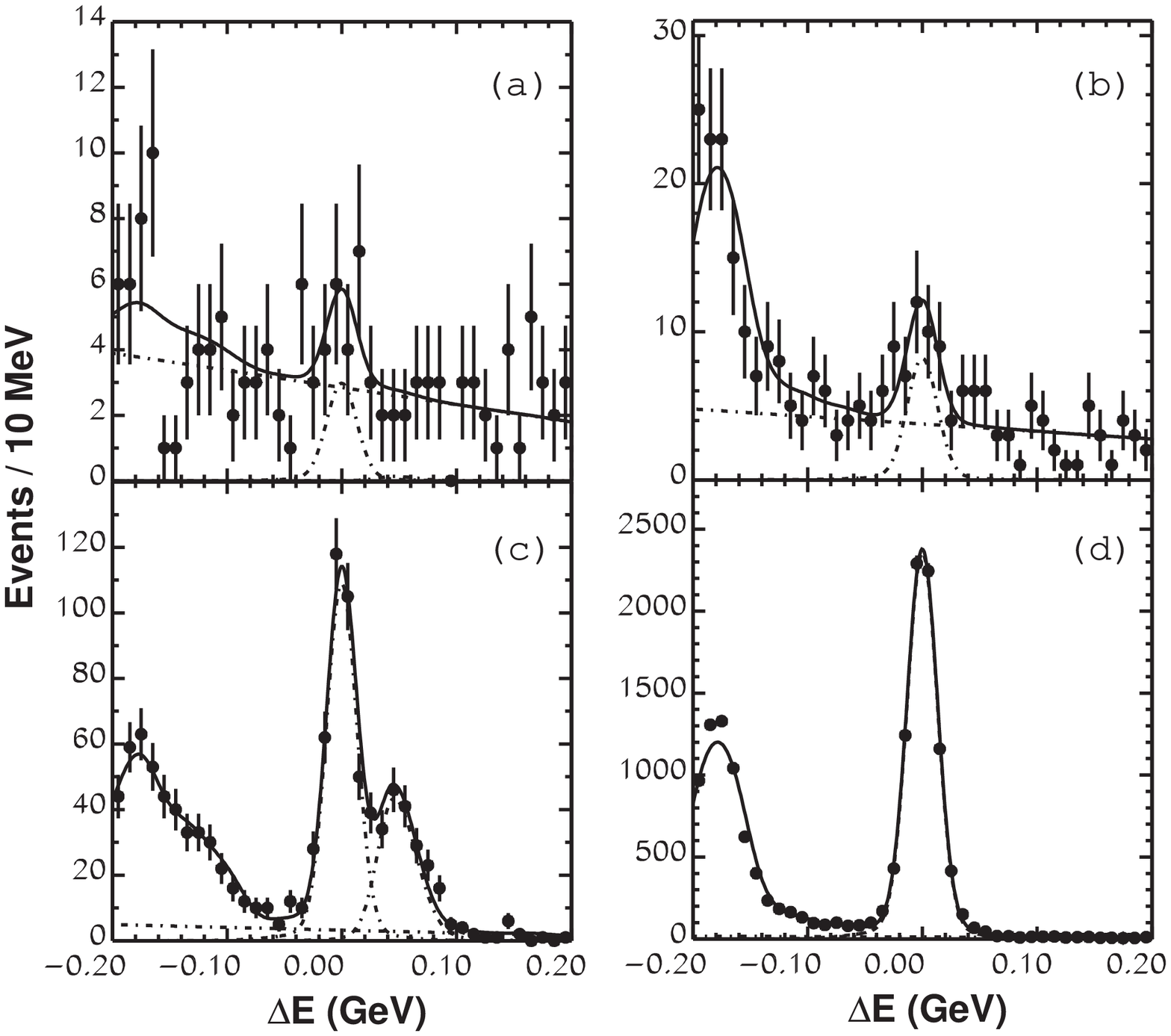}
    \caption{
      $\Delta E$ fit results for 
      (a) $B^- \to \dsup K^-$, 
      (b) $B^- \to \dsup \pi^-$, 
      (c) $B^- \to \dfav K^-$, and 
      (d) $B^- \to \dfav \pi^-$. 
      Charge conjugate modes are included in these plots.
    }
    \label{fig:fitting}
  \end{center}
\end{figure}

The ratio of branching fractions is defined as
\begin{equation} \nonumber 
  R_{Dh} \equiv 
  \frac{{\cal B}(B^-\to \dsup h^-)}{{\cal B}(B^-\to \dfav h^-)} = 
  \frac{N_{\dsup h^-}/\epsilon_{\dsup h^-}}{N_{\dfav h^-}/\epsilon_{\dfav h^-}}, 
\end{equation}
where $N_{\dsup h}$ ($N_{\dfav h}$) and  $\epsilon_{\dsup h^-}$ ($\epsilon_{\dfav h^-}$)
are the number of signal events and the reconstruction efficiency for the decay
$B^- \to \dsup h^-$ ($B^- \to \dfav h^-$),
and are given in Table~\ref{tab:yield}.

The ratios $R_{Dh}$ are calculated to be
\begin{eqnarray}
  R_{DK} &=& ( 2.3 \, ^{+1.6}_{-1.4} (\stat) \pm 0.1 (\syst) ) \times 10^{-2}, 
  \nonumber \\
  R_{D\pi} &=& ( 3.5 \, ^{+1.0}_{-0.9} (\stat) \pm 0.2 (\syst) ) \times 10^{-3}.
  \nonumber
\end{eqnarray}
Since the signal for $B^- \to \dsup K^-$ is not significant,
we set an upper limit at the $90\%$ confidence level (C.L.) of 
$R_{DK} < 4.4 \times 10^{-2}$.

The product branching fractions for $B^- \to \dsup h^-$ are determined as
\begin{equation} \nonumber
  {\cal B}(B^- \to \dsup h^-) = {\cal B}(B^- \to \dfav h^-) \times R_{Dh},
\end{equation}
and are given in Table~\ref{tab:yield}. 
A third uncertainty arises due to the error in the branching fraction
of $B^- \to \dfav h^-$, which is taken from~\cite{pdg}.
The uncertainties are statistics-dominated. 
For the $B^- \to \dsup K^-$ branching fraction, 
we set an upper limit at the $90\%$ C.L. of 
${\cal B}(B^- \to \dsup K^-) < 6.3 \times 10^{-7}$. 
For $B^- \to \dsup \pi^-$,
our measured branching fraction is consistent with expectation 
neglecting the contribution from $B^- \to \bar{D}^0 \pi^-$.

The ratio $R_{DK}$ is related to $\phi_3$ by
\begin{eqnarray}
  R_{DK} = \rb^2 + \rd^2 + 2 \rb \rd \cos \phi_3 \cos \delta, \nonumber
\end{eqnarray}
where~\cite{pdg}
\begin{eqnarray}
  \rb & \equiv & \left| \frac{A(B^- \to \bar{D}^0K^-)}{A(B^- \to D^0K^-)} \right|, 
  \:\:\:\:\: 
  \delta \equiv \deltab + \deltad, 
  \nonumber \\
  \rd & \equiv & \left| \frac{A(D^0 \to K^+\pi^-)}{A(D^0 \to K^-\pi^+)} \right| = 
  0.060 \pm 0.003, \nonumber
\end{eqnarray}
and $\deltab$ and $\deltad$ are the strong phase differences 
between the two $B$ and $D$ decay amplitudes, respectively. 
Using the above result, we obtain a limit on $\rb$. 
The least restrictive limit is obtained 
allowing $\pm 1\sigma$ variation on $\rd$ and assuming maximal interference
($\phi_3 = 0^\circ, \delta = 180^\circ$ or $\phi_3 = 180^\circ, \delta = 0^\circ$) 
and is found to be $\rb < \rblimit$.

\begin{figure}
  \begin{center}
    \includegraphics[width=0.5\textwidth]{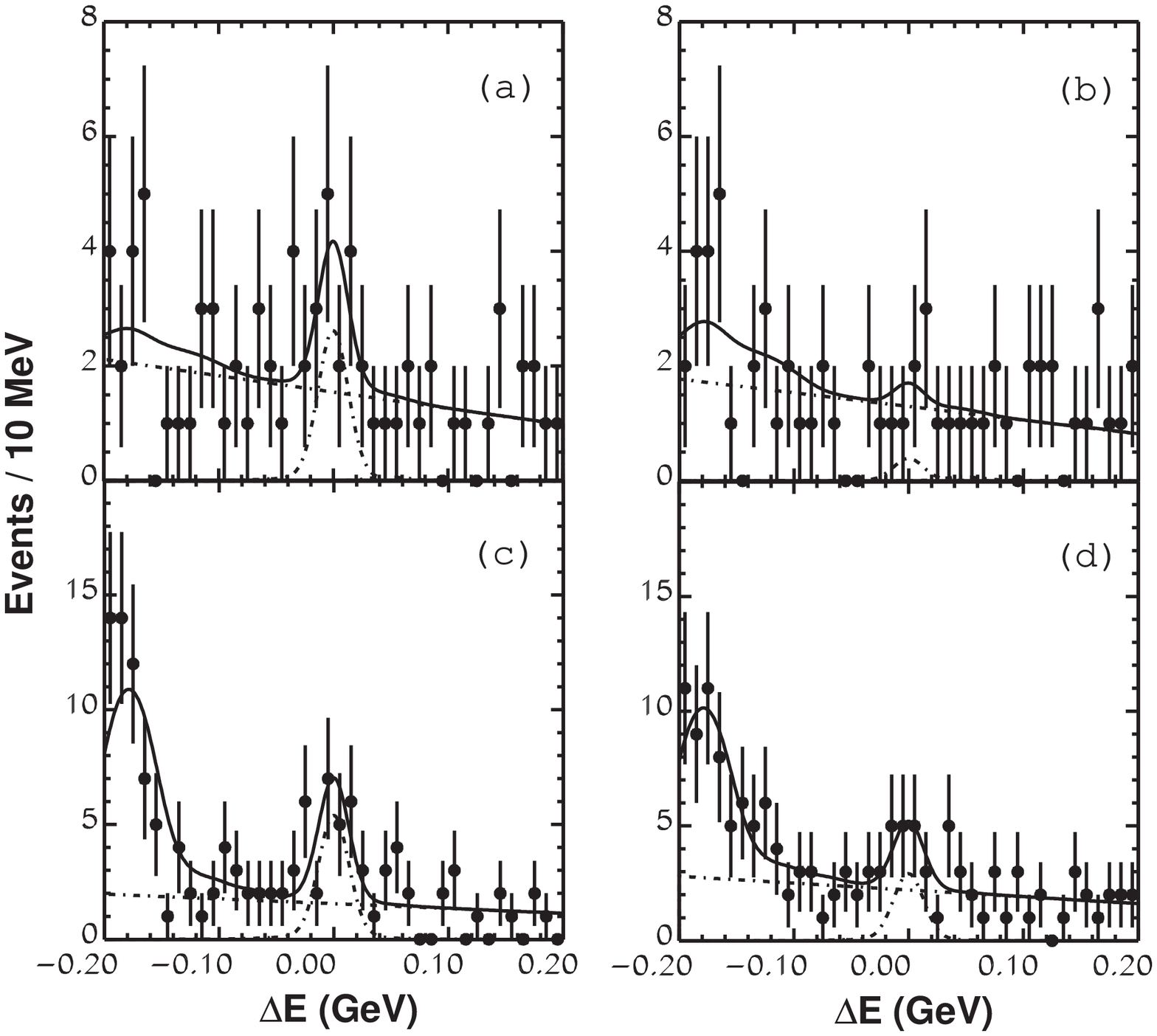}
    \caption{
      $\Delta E$ fit results for 
      (a) $B^- \to \dsup K^-$, 
      (b) $B^+ \to \dsup K^+$, 
      (c) $B^- \to \dsup \pi^-$, and 
      (d) $B^+ \to \dsup \pi^+$.
    }
    \label{fig:acp}
  \end{center}
\end{figure}

\begin{table*}
  \caption{
    Signal yields and partial rate asymmetries.
  }
\begin{tabular*}{\textwidth}{l@{\extracolsep{\fill}}c c c c c}
\hline\hline
      Mode & $N(B^-)$ & & $N(B^+)$ & & ${\cal A}_{Dh}$\\
      \hline
      $B \to \dsup K$ & $ 8.2\,^{+5.0}_{-4.3}$ & & $ 0.5\,^{+3.5}_{-2.8}$ & & $0.88\,^{+0.77}_{-0.62} \pm 0.06$ \\
      $B \to \dsup \pi$ & $18.8\,^{+6.3}_{-5.5}$ & & $10.1\,^{+5.5}_{-4.8}$ & & $0.30\,^{+0.29}_{-0.25} \pm 0.06$ \\
\hline\hline
    \end{tabular*}
  \label{tab:acp}
\end{table*}

We search for partial rate asymmetries 
${\cal A}_{Dh}$ in $B^\mp \to \dsup h^\mp$ decay, 
fitting the $B^-$ and $B^+$ yields separately for each mode, 
where ${\cal A}_{Dh}$ is determined as
\begin{equation}\nonumber 
  {\cal A}_{Dh} \equiv 
  \frac{
    {\cal B}(B^- \to \dsup h^-) - {\cal B}(B^+ \to \dsup h^+)
  }{
    {\cal B}(B^- \to \dsup h^-) + {\cal B}(B^+ \to \dsup h^+)
  }.
\end{equation}
The peaking background for $B^\mp \to \dsup K^\mp$ 
is subtracted assuming no $CP$ asymmetry. 
The fit results are shown in Fig.~\ref{fig:acp} and Table~\ref{tab:acp}. 
We find
\begin{eqnarray}
  {\cal A}_{DK} & = & 0.88 \, ^{+0.77}_{-0.62} (\stat) \pm 0.06 (\syst),
  \nonumber \\
  {\cal A}_{D\pi} & = & 0.30 \, ^{+0.29}_{-0.25} (\stat) \pm 0.06 (\syst). 
  \nonumber 
\end{eqnarray}

In summary, we observe $B^- \to \dsup \pi^-$ for the first time, 
with a significance of $\dpisig$. 
The size of the signal is consistent with 
expectation based on measured branching fractions~\cite{pdg}. 
The significance for $B^- \to \dsup K^-$ is $\dksig$ 
and we set an upper limit on the ratio of $B$ decay amplitudes 
$\rb < \rblimit$ at $90\%$ confidence level.

\section*{Acknowledgments}
We thank the KEKB group for the excellent operation of the
accelerator, the KEK cryogenics group for the efficient
operation of the solenoid, and the KEK computer group and
the NII for valuable computing and Super-SINET network
support.  We acknowledge support from MEXT and JSPS (Japan);
ARC and DEST (Australia); NSFC (contract No.~10175071,
China); DST (India); the BK21 program of MOEHRD and the CHEP
SRC program of KOSEF (Korea); KBN (contract No.~2P03B 01324,
Poland); MIST (Russia); MHEST (Slovenia);  SNSF (Switzerland); NSC and MOE
(Taiwan); and DOE (USA).

\end{document}